\documentclass{aa}  
\usepackage{float}
\usepackage{graphicx}
\catcode`\@=11
\def\gtsima{\ifmmode{\mathrel{\mathpalette\@versim>}}
    \else{$\mathrel{\mathpalette\@versim>}$}\fi}
\def\ltsima{\ifmmode{\mathrel{\mathpalette\@versim<}}
    \else{$\mathrel{\mathpalette\@versim<}$}\fi}
\def\@versim#1#2{\lower 2.9truept \vbox{\baselineskip 0pt \lineskip 
    0.5truept \ialign{$\m@th#1\hfil##\hfil$\crcr#2\crcr\sim\crcr}}}
\catcode`\@=12
\def\lae{Ly$\alpha$}
\def\ha{H$\alpha$}
\def\esc{$f_{esc}$(Ly$\alpha$)}
\def\ew{$EW_0$(Ly$\alpha$)}
\usepackage{txfonts}
\begin{document}
   \title{The VIMOS Ultra--Deep Survey (VUDS): fast increase in the
     fraction of strong Lyman-$\alpha$ emitters from z=2 to
     z=6\thanks{Based on data obtained with the European Southern
       Observatory Very Large Telescope, Paranal, Chile, under Large
       Program 185.A-0791.}}  \titlerunning{Ly-$\alpha$ fraction vs
     redshift}

   \author{P. Cassata \inst{1},\inst{2}
L. A. M. Tasca \inst{1},
O. Le~F\`evre\inst{1},
B. C. Lemaux\inst{1},
B. Garilli\inst{3},
V. Le Brun\inst{1},
D. Maccagni\inst{3},
L. Pentericci\inst{4},
R. Thomas\inst{1},
E. Vanzella\inst{4},
G. Zamorani \inst{4},
E. Zucca\inst{4},
R. Amorin\inst{5},
S. Bardelli\inst{4},
P. Capak\inst{13},
L. P. Cassar\`a\inst{3},
M. Castellano\inst{5},
A. Cimatti\inst{6},
J.G. Cuby\inst{1},
O. Cucciati\inst{6,4},
S. de la Torre\inst{1},
A. Durkalec\inst{1},
A. Fontana\inst{5},
M. Giavalisco\inst{14},
A. Grazian\inst{5},
N. P. Hathi\inst{1},
O. Ilbert\inst{1},
C. Moreau\inst{1},
S. Paltani\inst{10},
B. Ribeiro\inst{1},
M. Salvato\inst{15},
D. Schaerer\inst{11,9},
M. Scodeggio\inst{3},
V. Sommariva\inst{6,5},
M. Talia\inst{6},
Y. Taniguchi\inst{16},
L. Tresse\inst{1},
D. Vergani\inst{7,4},
P.W. Wang\inst{1},
S. Charlot\inst{8},
T. Contini\inst{9},
S. Fotopoulou\inst{10},
C. L\'opez-Sanjuan\inst{11},
Y. Mellier\inst{8},
N. Scoville\inst{13}
}
\authorrunning{P. Cassata et al.}  \offprints{P. Cassata}

   \institute{Aix Marseille Universit\'e, CNRS, LAM (Laboratoire
  d'Astrophysique de Marseille) UMR 7326, 13388, Marseille, France
              \email{paolo.cassata@lam.fr}
\and
Instituto de Fisica y Astronom\'ia, Facultad de Ciencias, Universidad de Valpara\'iso, Gran Breta\~{n}a 1111, Playa Ancha, Valparaíso, Chile
\and
INAF--IASF, via Bassini 15, I-20133,  Milano, Italy
\and
INAF--Osservatorio Astronomico di Bologna, via Ranzani,1, I-40127, Bologna, Italy
\and
INAF--Osservatorio Astronomico di Roma, via di Frascati 33, I-00040,  Monte Porzio Catone, Italy
\and
University of Bologna, Department of Physics and Astronomy (DIFA), V.le Berti Pichat, 6/2 - 40127, Bologna, Italy
\and
INAF--IASF Bologna, via Gobetti 101, I--40129,  Bologna, Italy
\and
Institut d'Astrophysique de Paris, UMR7095 CNRS,
Universit\'e Pierre et Marie Curie, 98 bis Boulevard Arago, 75014
Paris, France
\and
Institut de Recherche en Astrophysique et Plan\'etologie - IRAP, CNRS, Universit\'e de Toulouse, UPS-OMP, 14, avenue E. Belin, F31400
Toulouse, France
\and
Department of Astronomy, University of Geneva
ch. d'\'Ecogia 16, CH-1290 Versoix, Switzerland
\and
Geneva Observatory, University of Geneva, ch. des Maillettes 51, CH-1290 Versoix, Switzerland
\and
Centro de Estudios de F\'isica del Cosmos de Arag\'on, Teruel, Spain
\and
Department of Astronomy, California Institute of Technology, 1200 E. California Blvd., MC 249-17, Pasadena, CA 91125, USA
\and
Astronomy Department, University of Massachusetts, Amherst, MA 01003, USA
\and
Max-Planck-Institut f\"ur Extraterrestrische Physik, Postfach 1312, D-85741, Garching bei M\"unchen, Germany
\and
Research Center for Space and Cosmic Evolution, Ehime University, Bunkyo-cho 2-5, Matsuyama 790-8577, Japan}

   \date{Received .....; accepted .....}

  \abstract 
{} 
  {The aim of this work is to constrain the evolution of the fraction
    of strong \lae~emitters among UV selected star--forming galaxies
    at $2<z<6$, and to measure the stellar escape fraction of \lae~
    photons over the same redshift range.}
  {We exploit the ultradeep spectroscopic observations with VIMOS on
    the VLT collected by the VIMOS Ultra--Deep Survey (VUDS) to build
    an unique, complete, and unbiased sample of $\sim4000$
    spectroscopically confirmed star--forming galaxies at $2<z<6$. Our
    galaxy sample includes UV luminosities brighter than $M^*_{FUV}$
    at $2<z<6$, and luminosities down to one magnitude fainter than
    $M^*_{FUV}$ at $2<z<3.5$. }
  {We find that 80\% of the star--forming galaxies in our sample have
    $EW_0(Ly\alpha)<10$\AA, and correspondingly \esc$<1$\%. By
    comparing these results with the literature, we conclude that the
    bulk of the \lae~luminosity at $2<z<6$ comes from galaxies that
    are fainter in the UV than those we sample in this work. The
    strong \lae~emitters constitute, at each redshift, the tail of the
    distribution of the galaxies with extreme \ew~and \esc. This tail
    of large \ew~and \esc~becomes more important as the redshift
    increases, and causes the fraction of strong \lae~with
    \ew$>25$\AA~to increase from $\sim$5\% at $z\sim2$ to $\sim$30\%
    at $z\sim6$, with the increase being stronger beyond z$\sim4$. We
    observe no difference, for the narrow range of UV luminosities
    explored in this work, between the fraction of strong
    \lae~emitters among galaxies fainter or brighter than $M^*_{FUV}$,
    although the fraction for the faint galaxies evolves faster, at
    $2<z<3.5$, than for the bright ones. We do observe an
    anticorrelation between E(B-V) and \esc: generally galaxies with
    high \esc~also have small amounts of dust (and
    vice versa). However, when the dust content is low (E(B-V)$<$0.05)
    we observe a very broad range of \esc, ranging from 10$^{-3}$ to
    1. This implies that the dust alone is not the only regulator of
    the amount of escaping \lae~photons.}
{}

   \keywords{Cosmology: observations -- Galaxies: fundamental
     parameters -- Galaxies: evolution -- Galaxies: formation }

   \maketitle
%

\section{Introduction}

Narrowband surveys targeting the strong \lae~emission from
star--forming galaxies (Lyman-$\alpha$ emitters, LAEs;
Partridge\&Peebles~1967; Djorgovski~et~al.~1985; Cowie~\&Hu~1998;
Hu~et~al.~2004; Kashikawa~et~al.~2006; Gronwall~et~al.~2007;
Murayama~et~al.~2007; Ouchi~et~al.~2008; Nilsson~et~al.~2009) and
broadband surveys targeting the deep Lyman break (LBG;
Steidel~et~al.~1999; Bouwens~\&~Illingworth~2006; Bouwens~et~al.~2010;
McLure~et~al.~2011) have been very successful at exploring the
high--redshift Universe.  However, the overlap between the populations
selected by the two techniques is still debated: LAEs are claimed to
be forming stars at rates of $1\div10 M_{\odot}yr^{-1}$
(Cowie~\&~Hu~1998; Gawiser~et~al.~2006; Pirzkal~et~al.~2007), to have
stellar masses on the order of $10^8\div10^9M_{\odot}$ and to have
ages smaller than 50 Myr (Pirzkal~et~al.~2007; Gawiser~et~al.~2007;
Nilsson~et~al.~2009), while LBGs have in general a broader range of
properties (Reddy~et~al.~2006; Hathi~et~al.~2012;
Schaerer,~de~Barros~\&~Stark~2011; but see also Kornei~et~al.~2010).

Steidel~et~al.~(2000) and Shapley~et~al.~(2003) showed that only
$\sim20$\% of z$\sim3$ LBGs have a \lae~emission strong enough to be
detected with the narrowband technique. Recently, many authors have
investigated the evolution with the redshift of the fraction of strong
\lae~emitters among LBG galaxies. Stark~et~al.~(2010;~2011) showed
that this fraction evolves with redshift, and that the overall
fraction is smaller (and that the rate of evolution is slower) for UV
bright galaxies ($-21.75<M_{UV}<-20.25$) than for UV faint
($-20.25<M_{UV}<-18.75$) galaxies; they find that the fraction of UV
faint galaxies with strong (\ew$>25$~\AA) \lae~emission is around 20\%
at $z\sim2\div3$ and reaches $\sim50\div60$\% at $z\sim6$. At higher
redshift ($z>6\div8$), many authors claim a sudden drop in the
fraction of spectroscopically confirmed LBGs with strong \lae~emission
(Fontana~et~al.~2010; Pentericci~et~al.~2011; Ono~et~al.~2012;
Schenker~et~al.~2012; Caruana~et~al.~2014), interpreting this as the
observational signature of the increasing fraction of netural hydrogen
between $z\sim6$ and $z\sim7$ due to the tail end of the reionization,
although Dijkstra~et~al.~(2014) has argued that the effect can be due
to a variation of the average escape fraction over the same redshift
range.

However, the bulk of studies of the \lae~fraction at $3<z<8$
(Stark~et~al.~2010;~2011; Pentericci~et~al.~2011) are based on a
hybrid photometric-spectroscopic technique: the denominator of the
fraction (i.e., the {\it total} number of star--forming galaxies at
those redshifts) is only constrained by photometry, and thus its
determination relies on the strong assumption that the contamination
by low--$z$ interlopers and incompleteness are fully understood and
well controlled. The numerator of the fraction is the number of the
LBGs that are observed with spectroscopy, and for which a strong \lae~
rest--frame Equivalent Width (EW$_0>$25~\AA) is measured. In fact, the
LBGs for which this experiment is done have a UV continuum that is
generally too faint to be detected, even with the most powerful
spetrographs on 10-meter class telescopes.  Recently,
Mallery~et~al.~(2012) combined a sample of LAEs and LBGs to constrain
the evolution of this fraction, confirming earlier results by
Stark~et~al.~(2010;~2011). Given the nature of the selection of these
samples, it is important to make a robust estimate of the evolution of
the \lae~fraction covering as wide a range in redshift as possible,
and based on larger samples.

The \lae~is interesting not only because it allows for the exploration
of the high--redshift universe. In fact, its observed properties can
give a lot of information about the physical condition of
star--forming galaxies.  \lae~is thought to be mainly produced by star
formation, as the contribution of AGN activity to the \lae~population
at $z<4$ is found to be less than 5\% (Gawiser~et~al.~2006;
Ouchi~et~al.~2008; Nilsson~et~al.~2009; Hayes~et~al.~2010). Because of
its resonant nature, \lae~photons are easily scattered, shifted in
frequency, and absorbed by the neutral hydrogen and/or by the dust. As
a result, in general, \lae~emission is more attenuated than other UV
photons, with the \lae~escape fraction (i.e., the fraction of the
\lae~ photons that escape the galaxies) that depends strongly on the
relative kinematics of the HII and HI regions, dust content, and
geometry (Giavalisco~et~al.~1996; Kunth~et~al.~1998;
Mas-Hesse~et~al.~2003; Deharveng~et~al.~2008; Hayes~et~al.~2014).  As
a result of their nature, \lae~photons are found to be scattered at
much larger scales than UV photons (Steidel~et~al.~2011;
Momose~et~al.~2014).

Predicting the escape fraction of the \lae~photons as a function of
the galaxy properties involves including all the complex effects of
radiative transfer of such photons. Developing the first models by
Charlot~\&~Fall~(1993), Verhamme~et~al.~(2006; 2008; 2012) and
Dijkstra~et~al.~(2006; 2012) made huge progress in predicting the
shape of the \lae~emission as a function of the properties of the ISM,
the presence of inflows/outflows, and dust. Verhamme~et~al.~(2006;
2008) predicted a correlation between \esc~and E(B-V), with the escape
fraction being higher in galaxies with low dust
content. Verhamme~et~al.~(2012) and Dijkstra~et~al.~(2012) studied the
escape fraction of \lae~photons through a 3D clumpy medium,
constraining the dependence on the column density of neutral hydrogen
and on the viewing angle.

A lot of effort has been recently put to constrain the correlation
between the \lae~properties and the general properties of
star--forming galaxies (e.g., dust attenuation, SFR, stellar mass) in
the local Universe.  Hayes~et~al.~(2014) and Atek~et~al.~(2014) have
found that \lae~photons escape more easily from galaxies with low dust
content. At high redshift, although on samples that are much smaller
than the one we use in this paper, a similar trend has been found by
Kornei~et~al.~(2010) and Mallery~et~al.~(2012), respectively at
$z\sim3$ and at $4<z<6$. In this paper, we look for this correlation
using a sample that is respectively five and ten times larger than the
ones used by Mallery and Kornei.

The aim of this paper is to estimate the evolution of the fraction of
strong \lae~emitters as a function of the redshift, exploiting data
from the new VIMOS Ultra--Deep Survey (VUDS). The goal is twofold:
first, to put on firmer grounds the trends that have been found with
photometric LBG samples (Stark~et~al.~2010; 2011) and improve on the
knowledge of the evolution of the \lae~fraction; second, to offer the
theoreticians a reference sample of galaxies with robust spectroscopic
redshifts, with a well measured \ew~distribution. In fact, in this
paper, we select a sample of galaxies, sliced in volume limited
samples according to different recipes, for which we have a
spectroscopic redshift in $\sim$90\% of the cases. The continuum is
detected for almost all objects in the sample, thus allowing a robust
measurement of the redshift based on the UV absorption features even
in absence of \lae.
\begin{figure*}[!ht]
  \centering \includegraphics[width=.85\textwidth]{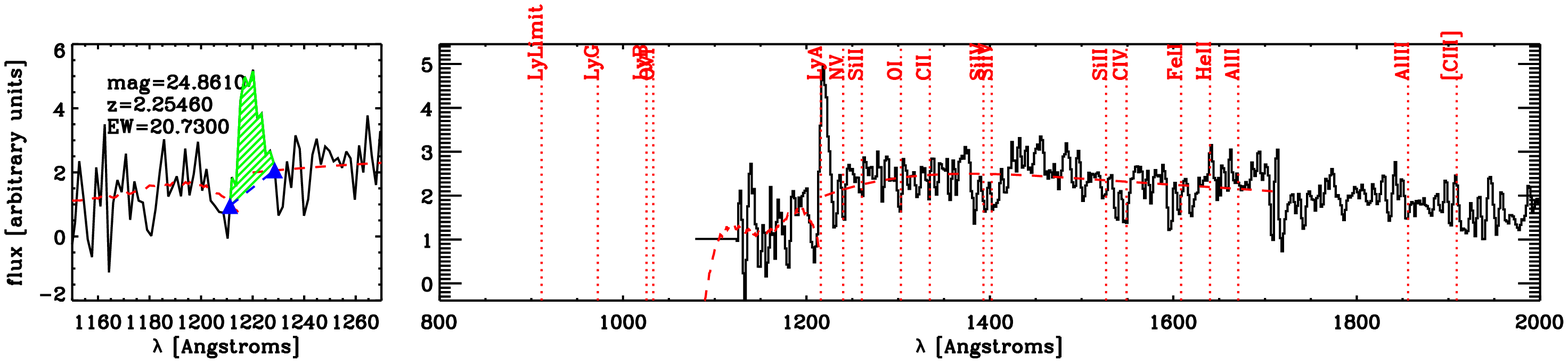}
  \centering \includegraphics[width=.85\textwidth]{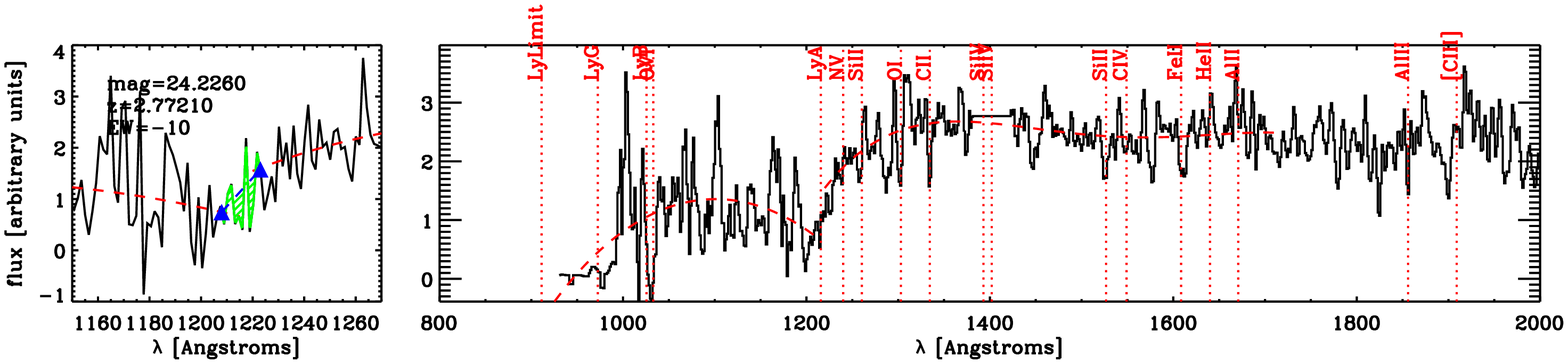}
  \centering \includegraphics[width=.85\textwidth]{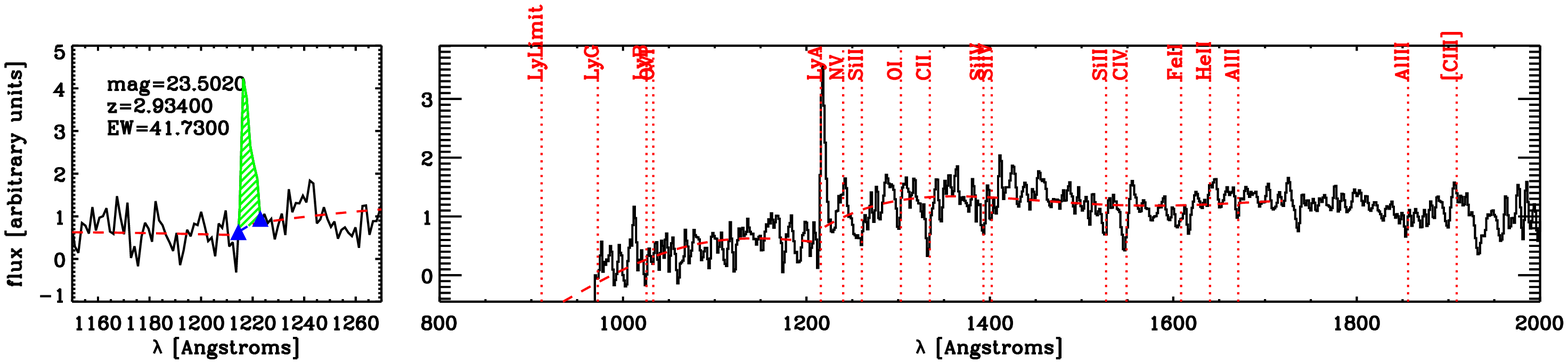}
  \centering \includegraphics[width=.85\textwidth]{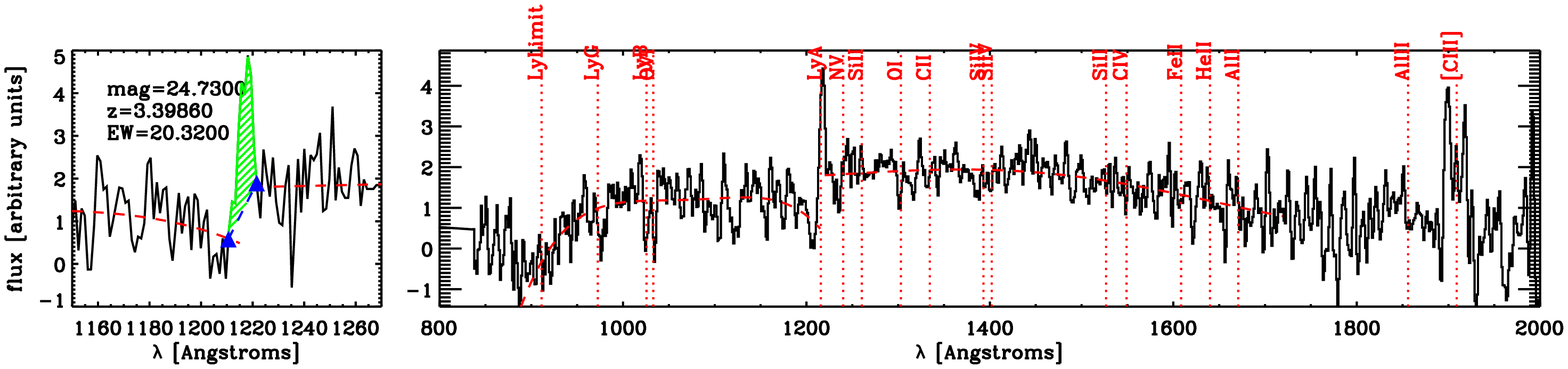}
  \centering \includegraphics[width=.85\textwidth]{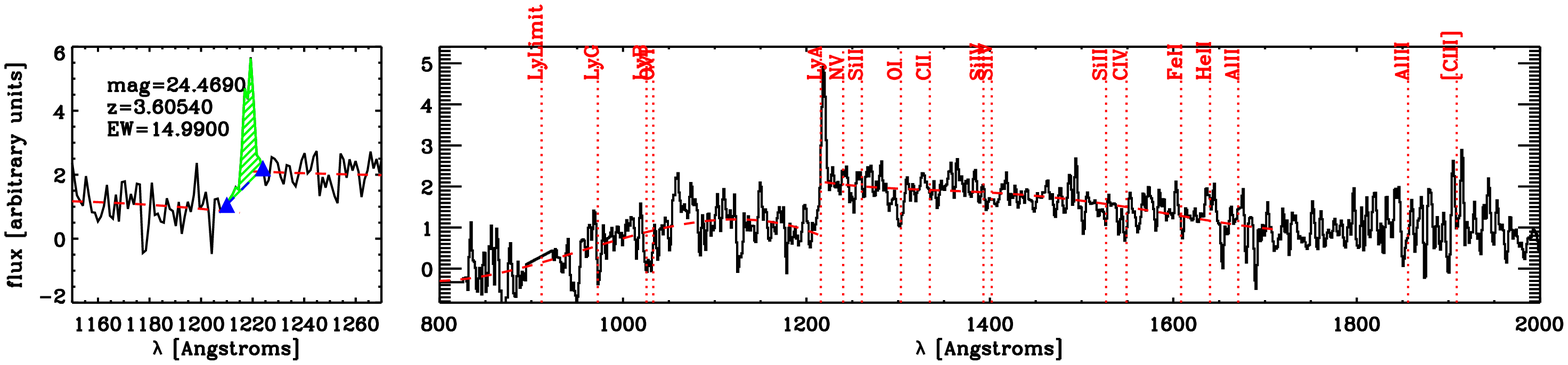}
  \centering \includegraphics[width=.85\textwidth]{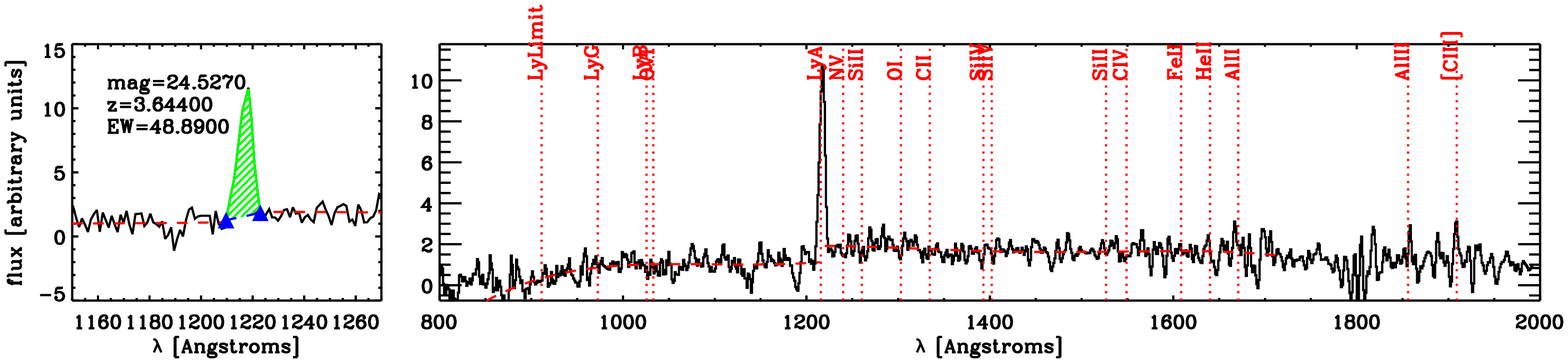}
  \caption{Six examples of spectra for the galaxies in the sample. The
    left panels show the region around \lae, while the right ones show
    the full spectrum, with the most common UV rest-frame lines
    highlighted in red. These examples are chosen to be representative
    of the $i-$band magnitudes, redshifts and \lae~equivalent widths
    covered by the sample presented in this work. The red dashed
    curves show polynomial fits to the continuum: for each spectrum,
    the region between 912~\AA~and \lae~and the region between \lae~
    and 2000~\AA~are fitted separately. We note that the fits are not
    used at all in the analysis presented in this paper; they only
    provide a guidance to assess the continuum around \lae. The blue
    triangles show the points on the continuum bracketing the \lae~
    line, shown in green.}
  \label{ex}%
\end{figure*}

\begin{table}
 \centering
\begin{tabular}
{clr@{$\pm$}rr@{$\pm$}lr@{$\pm$}lr@{$\pm$}lr@{$\pm$}lc}

\hline\hline \noalign{\smallskip} 

\multicolumn{1}{c}{} & \multicolumn{2}{c}{$2<z<2.7$} & \multicolumn{2}{c}{$2.7<z<3.5$} & \multicolumn{2}{c}{$3.5<z<4.5$} & \multicolumn{2}{c}{$4.5<z<6$}\\ 

\noalign{\smallskip} \hline
\noalign{\smallskip}

\multicolumn{1}{c} {f=0}& \multicolumn{2}{c}{83(0)} & \multicolumn{2}{c}{90(0)} & \multicolumn{2}{c}{40(0)} & \multicolumn{2}{c}{18(0)}\\ 
\noalign{\smallskip}
\multicolumn{1}{c} {f=1}& \multicolumn{2}{c}{299(2)} & \multicolumn{2}{c}{200(3)} & \multicolumn{2}{c}{88(2)} & \multicolumn{2}{c}{14(0)}\\ 
\noalign{\smallskip}
\multicolumn{1}{c} {f=2}& \multicolumn{2}{c}{614(24)} & \multicolumn{2}{c}{614(17)} & \multicolumn{2}{c}{163(5)} & \multicolumn{2}{c}{47(3)}\\ 
\noalign{\smallskip}
\multicolumn{1}{c} {f=3,4}& \multicolumn{2}{c}{646(106)} & \multicolumn{2}{c}{701(153)} & \multicolumn{2}{c}{205(57)} & \multicolumn{2}{c}{41(22)}\\ 
\noalign{\smallskip}
\multicolumn{1}{c} {f=9}& \multicolumn{2}{c}{28(15)} & \multicolumn{2}{c}{31(10)} & \multicolumn{2}{c}{19(6)} & \multicolumn{2}{c}{20(5)}\\ 

\noalign{\smallskip} \hline
\noalign{\smallskip}
\end{tabular}
\caption{The final sample of galaxies used in this work, divided in 4
  redshift bins, as a function of the spectroscopic quality flag.  The
  number in parentheses indicates the number of objects at that
  redshift and of that spectroscopic quality flag that have
  EW$_0>25$~\AA. }\label{tab:sample}
\end{table}

Our selection is not based on LBG or narroband techniques, that are
prone to incompleteness and contamination, but it is rather based on
the magnitude in the $i'-$band and on the photometric redshifts
measured on the full Spectral Energy Distribution (SED) of
galaxies. The most important point to emphasize is that our flux
selection is completely independent of the presence of \lae, at least
up to $z\sim5$, because it enters the photometric $i'-$band only at
$z>5$: since the $i'-$band does not contain the \lae~line, objects
with strong \lae~emission have not a boosted $i'-$band
magnitude. Moreover, when the photo-$z$ are computed, some variable
\lae~flux (as for other lines like OII, OIII and H$\alpha$) is added
to the SED: this ensures that even objects with large \lae~flux are
reproduced by the template set that is used to compute the
photo-z. This also implies that if our selection is incomplete at some
redshift, the incompleteness is also independent of the presence (or
absence) of \lae.

For these reasons, this sample is ideal to study the \lae~properties
of a well controlled sample of star--forming galaxies. The fraction of
strong \lae~emitters among star--forming galaxies is completely
constrained by spectroscopy, as is also the case for non-\lae~
emitters.

Throughout the paper, we use a standard Cosmology with $\Omega_M=0.3$,
$\Omega_{\Lambda}=0.7$ and $h=0.7$. Magnitudes are in the AB system.



\section{Data}
The data used in this study are drawn from the VIMOS Ultra--Deep
Survey (VUDS), an ESO large program with the aim of collecting spectra
and redshifts for around 10,000 galaxies to study early phases of
galaxy formation at $2<z<6$. To minimize the effect of cosmic
variance, the targets are selected in three independent extragalactic
fields: COSMOS (Scoville~et~al.~2007), the CFHTLS-D1 Field
(Cuillandre~et~al.~2012) and the Extended-Chandra-Deep-Field (ECDFS;
e.g., see Cardamone~et~al.~2010). The survey is fully presented in
Le~F\`evre~et~al.~(2014).

\subsection{Photometry}

\begin{figure}[!ht]
  \centering \includegraphics[width=\columnwidth]{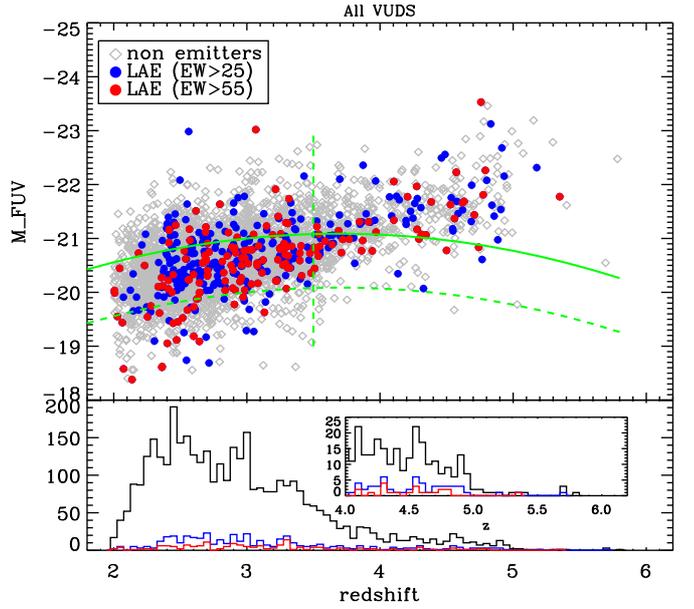}
  \caption{{\it Top panel:} Absolute magnitude in the far-UV band as a
    function of the redshift, for all VUDS galaxies at $2<z<6$ (gray
    diamonds), for the galaxies with EW$_0>25$~\AA~(blue circles) and
    for the galaxies with EW$_0>55$~\AA~(red circles).  The green
    continuous curve indicates the evolving $M^*$ as a function of the
    redshift as derived from the compilation by Hathi~et~al.~2010; the
    dashed green curve indicates $M^*+1$.  The vertical dashed line
    shows z=3.5, the redshift up to which the faint sample is
    complete. {\it Bottom panel:} Redshift distribution of the all the
    VUDS galaxies at $2<z<6$ (black line) and of the VUDS galaxies
    with EW$_0>25$\AA~ (blue histogram) and EW$_0>55$\AA~ (red
    histogram).}
  \label{fuv}%
\end{figure}

The three extragalactic fields targeted by the VUDS survey are three
of the most studied regions of the sky, and they have been imaged by
some of the most powerful telescopes on earth and in the space,
including CFHT, Subaru, HST and Spitzer. For more details, we refer
the reader to Le~F\`evre~et~al.~(2014), where more detailed
information can be found.

The COSMOS field was observed with HST/ACS in the F814W filter
(Koekemoer~et~al.~2007). Ground based imaging includes deep
observations in $g',r',i'$ and $z'$ bands from the Subaru SuprimeCam
(Taniguchi~et~al.~2007) and $u^*$ band observations from CFHT Megacam
from the CFHT-Legacy Survey. Moreover, the UltraVista survey is
acquiring very deep near-infrared imaging in the Y, J, H and K bands
using the VIRCAM camera on the VISTA telescope
(McCracken~et~al.~2012), and deep Spitzer/IRAC observations are
available (Sanders~et~al.~2007; Capak~et~al.~in~prep.). The CANDELS
survey (Grogin~et~al.~2012) also provided WFC3 NIR photometry in the
F125W and F160W bands, for the central part of the COSMOS field.

The ECDFS field is covered with deep UBVRI imaging down to
$R_{AB}=25.3$ (5$\sigma$, Cardamone~et~al.~2010 and references
therein). For the central part of the field, covering
$\sim160$arcmin$^2$, observations with HST/ACS in the F435W, F606W,
F775W and F850LP are available (Giavalisco~et~al.~2004), together with
the recent CANDELS observations in the J, H and K bands.  The SERVS
Spitzer-warm obtained 3.6$\mu$m and 4.5$\mu$m (Mauduit~et~al.~2012)
that complement those obtained by the GOODS team at 3.6$\mu$m,
4.5$\mu$m, 5.6$\mu$m and 8.0$\mu$m.

The VVDS-02h field is observed in the BVRI at the CFHT
(Le~F\`evre~et~al.~2004), and later received deeper observations in
the $u^*, g', r'$ and $i'$ bands as part of the CFHTLS survey
(Cuillandre~et~al.~2012). Deep infrared imaging has been obtained with
the WIRCAM at CFHT in YJHK bands down to $K_s$=24.8
(Bielby~et~al.~2012). This field was observed in all Spitzer bands as
part of the SWIRE survey (Lonsdale~et~al.~2003), and recently deeper
data were obtained as part of the SERVS survey (Mauduit~et~al.~2012).

\subsection{Target selection}
The aim of the VUDS survey is to build a well controlled and complete
spectroscopic sample of galaxies in the redshift range $2\lesssim
z\lesssim 6$. To achieve this goal, with the aim of being as inclusive
as possible, we combined different selection criteria such as
photometric redshifts, color-color and narrow-band selections. All the
details of the selection can be found in Le~F\`evre~et~al.~(2014).

For this paper, we limited the analysis to the objects selected by the
primary selection, that is based on photometric redshift and magnitude
in the $i'-$band. In particular, only galaxies with auto magnitude in
the $i'-$band $22.5<m_i<25$ are included. If an object has a
photometric redshift $z_p>2.4-\sigma_{z_p}$ (where $\sigma_{z_p}$
denotes the 1-$\sigma$ error on the photometric redshift) or if the
second peak of the photometric redshift Probability Distribution
Function (zPDF) $z_{p,2}>2.4$, this object is included in the target
list.

\subsection{Spectroscopy}

\begin{figure}
  \centering \includegraphics[width=\columnwidth]{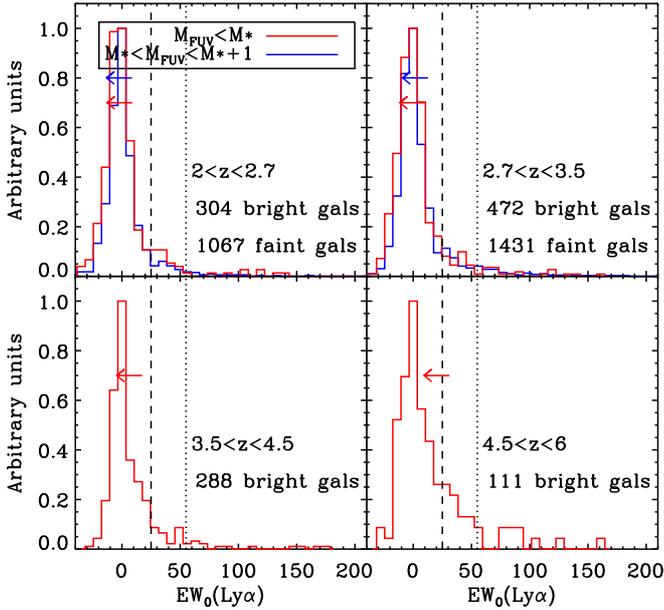}
  \caption{Rest-frame equivalent width EW$_0$ of the \lae~line in four
    redshift bins. The dashed and dotted lines, respectively at
    EW$_0$=25 and 55~\AA, represent the two thresholds that we apply
    in the analysis. The red and blue histogram indicate the bright
    sample ($M_{FUV}<M^*$) and the faint ($M^*<M_{FUV}<M^*+1$) one,
    respectively. 80\% of the galaxies in each panel have an \ew~below
    the value indicated by the arrow.}
  \label{ew_dist}%
\end{figure}

The spectroscopic observations were carried out with the VIMOS
instrument on the VLT. A total of 640h were allocated, including
overheads, starting in periods P85 and ending in P93 (end of 2013) to
observe a total of 16 VIMOS pointings. The spectroscopic MOS masks
were designed using the $vmmps$ tool (Bottini~et~al.~2005) to maximize
the number of spectroscopic targets that could be placed in them. In
the end, around 150 targets were placed in each of the 4 VIMOS
quadrants, corresponding to about 600 targets per pointing and
about 9000 targets in the whole survey. The same spectroscopic mask
was observed once for 14h with the LRBLUE grism (R=180) and for 14h
with the LRRED grism (R=210), resulting in a continuous spectral
coverage between $\lambda=3650$~\AA~and
$\lambda=9350$~\AA. Le~F\`evre~et~al.~(2013) used the data from the
VVDS survey to estimate the redshift accuracy of this configuration,
constraining it to $\sigma_{z_{spec}}=0.0005(1+z_{spec})$,
which corresponds to $\sim150km/s$.

The spectroscopic observations are reduced using the VIPGI code
(Scodeggio~et~al.~2005). First, the individual 2D spectrograms coming
from the 13 observing batches (OBs), in which the observations are
splitted, are extracted. Sky subtraction is performed with a low order
spline fit along the slit at each wavelength sampled. The sky
subtracted 2D spectrograms are combined with sigma clipping to produce
a single stacked 2D spectrogram calibrated in wavelength and
flux. Then, the objects are identified by collapsing the 2D
spectrograms along the dispersion direction. The spectral trace of the
target and other detected objects in a given slit are linked to the
astrometric frame to identify the corresponding target in the parent
photometric catalogue. At the end of this process 1D sky-corrected,
stacked and calibrated spectra are extracted. For more detail, we
refer the reader to Le~F\`evre~et~al.~(2014).

The redshift determination procedure follows the one that was
optimized for the VVDS survey (Le~F\`evre~et~al.~2005), later used in
the context of the zCOSMOS survey (Lilly~et~al.~2007) and VIPERS
(Guzzo~et~al.~2014): each spectrum is analyzed by two different team
members; the two independent measurements are then reconciled and a
final redshift with a quality flag are assigned. The EZ tool
(Garilli~et~al.~2010), a cross-correlation engine to compare spectra
and a wide library of galaxy and star templates, is run on all objects
to obtain a first guess of the redshift; after a visual inspection of
the solutions, it is run in manual mode to refine them, if necessary.

A quality flag is assigned to each redshift, repeating the same scheme
already used for the VVDS, COSMOS and VIPERS survey. The flag scheme
was thoroughly tested in the context of the VVDS survey, on spectra of
similar quality than the one we have for VUDS, and it is remarkably
stable, since the individual differences are smoothed out by the
process that involves many people (Le F\`evre~et~al.~2013). In
particular, Le F\`evre~et~al.~(2013) estimated the reliability of each
class:\\ \\ - Flag 4: 100\% probability to be correct\\ - Flag 3:
95--100\% probability to be correct\\ - Flag 2: 75--85\% probability
to be correct\\ - Flag 1: 50--75\% probability to be correct\\ - Flag
0: no redshift could be assigned\\ - Flag 9: the spectrum has a single
emission line.\\

The equivalent width (EW) of the \lae~line was measured manually using
the {\it splot} tool in the {\it noao.onedspec} package in IRAF,
similarly to Tresse~et~al.~(1999). We first put each galaxy spectrum
in its rest--frame according to the spectroscopic redshift. Then, two
continuum points bracketing the \lae~are manually marked and the
rest--frame equivalent width is measured. The line is not fitted with
a Gaussian, but the flux in the line is obtained integrating the area
encompassed by the line and the continuum. This method allows the
measurement of lines with asymmetric shapes (i.e., with deviations from
Gaussian profiles), which is expected to be the case for most \lae~
lines.  The interactive method also allows us to control by eye the
level of the continuum, taking into account defects that may be
present around the line measured. It does not have the objectivity of
automatic measurements, but, given the sometimes complex blend between
\lae~emission and \lae~absorption, it does produce reliable and
accurate measurements. We stress here that the $m_i<25$ selection
ensures that the continuum around \lae~is well detected for all
galaxies in our sample, even for galaxies with spectroscopic flag 1
(the lowest quality) and 9 (objects with a single emission line).

We report in Figure~\ref{ex} six examples of spectra in our sample.
These objects are representative of the range of magnitudes, redshifts
and \lae~EW covered by the sample. We note that the fit to the continuum
shown in the examples is not used at all for the scientific analysis
presented in this paper. It is only shown as a guide to select the
continuum points bracketing the \lae~line. For some more examples of
spectra used in this study, we refer the reader to Le~F\`evre et
al. (2014).

\subsection{Absolute magnitudes and masses}

We fitted the spectral energy distributions of galaxies in the survey
using the Le~Phare tool (Ilbert~et~al.~2006), following the same
procedure described in Ilbert~et~al.~(2013). The redshift is fixed to
the spectroscopic one for objects with flags 1, 2, 3, 4 and 9. It is
fixed to the photometric one for objects with spectroscopic flag 0. In
particular, we used the suite of templates by
Bruzual~\&~Charlot~(2003) with 3 metallicities ($Z=0.004, 0.008,
0.02$), assuming the Calzetti~et~al.~(2000) extinction curve. We used
exponentially declining star formation histories, with nine possible
$\tau$ values ranging from 0.1 Gyr (almost istantaneous burst) to 30
Gyr (smooth and continuous star formation). Moreover, since there is
now growing evidence that exponentially increasing models can better
describe the SFH of some galaxies beyond $z=2$ (Maraston~et~al.~2010;
Papovich~et~al.~2011; Reddy~et~al.~2012), we also included two delayed
SFH models, for which SFR$\propto$$\tau^{-2}t e^{-t/\tau}$, with
$\tau$ that can be 1 or 3 Gyr. For all models, the age ranges 0.05 Gyr
and the age of the Universe at the reshift of each galaxy. Since all
galaxies in our sample have $z>2$, and the Universe was already 3 Gyr
old at that redshift, all the galaxies in our sample fitted with the
delayed SFH with $\tau=3$ Gyr are caught when the SFH is rising.  In
the end, we let the fitting procedure decide what the best model is
for each galaxy, if increasing or declining.

For each synthetic SED, emission lines are added to the synthetic
spectra, with their luminosity set by the intensity of the SFR. Once
the \ha~luminosity is obtained from the SFR applying the classical
Kennicutt~(1998) relations, the theoretical \lae~luminosity is
obtained assuming case B recombination (Brocklehurst~1971). Then, the
actual \lae~luminosity that is added to the SED is allowed to vary
between half and double the theoretical value. The absolute magnitudes
are then derived by convolving the best template with the filter
responses. The output of this fitting procedure also includes the
stellar masses, star formation rates and extinction E(B-V).
\subsection{The dataset}\label{dataset}
For this paper we limit the analysis to the redshift range $2<z<6$.
The lower limit is the lowest redshift for which the \lae~line is
redshifted into our spectral coverage. For the upper limit, in theory,
we could detect \lae~in emission up to $z\sim6.5$, but the scarcity of
objects at $z>6$ in the VUDS survey forced us to limit the analysis to
$z\sim6$.  We limit the analysis to the galaxies with $m_i<25$: at
these magnitudes the continuum is always detected with signal-to-noise
ratio per resolution element S/N$\sim10$, ensuring that UV emission
and absorption lines with intrinsic $|EW|\gtrsim2$ are easily
identified in the spectra and the redshift determination is quite
reliable, for both spectra with \lae~in emission and absorption. As we
already said in the Introduction, the \lae~line enters the $i'-$band
only at $z>5$: this ensures that no detection bias affects our
analysis at $z<5$. We have in our sample only 12 galaxies at $5<z<6$:
we choose to keep them for our analysis throughout the paper, but we
will be extremely cautious to draw strong conclusions for that
redshift range.

We also include in the analysis secondary objects, that is the objects
that serendipitously fall in the spectroscopic slit centered on a
target, for which a spectrum is obtained in addition to that of the
target. For these objects, if they are brighter than $m_i<25$, a
spectroscopic redshift can also be easily assigned. However, only 2\% of
the final sample is made by secondary objects, that in any case only
marginally affect the main result of this paper. The database contain
4420 objects with $m_i<25$ that have been targeted by spectroscopy. Of
these, 3129 have a high reliability spectroscopic redshift in the
range $2<z<6$, with a spectroscopic flag 2, 3 or 4. Of the remaining
objects, 1058 have a more uncertain spectroscopic redshift, with a
quality flag 1: statistically, Le~F\`evre~et~al.~(2013) showed that
they are right in 50--75\% of the cases. For the purpose of this
paper, we decided to trust their spectroscopic redshift if the
difference between the photometric and spectroscopic redshift is
smaller than 10\%; otherwise, we fix the redshift to the photometric
one. We stress that almost all of the 1058 objects do not show any
strong emission line in their spectra that could be interpreted as
\lae, and that could help to assign a reliable spectroscopic
redshift. Thus, they will not be part of the sample of strong
\lae~emitters, but they will contribute to the total sample of
galaxies without \lae~emission, hence setting a lower limit to the
\lae~ fraction (see below). In the end, only 601 of these 1058 objects
with spectroscopic flag 1 survive the check against the photometric
redshift ($\sim60$\%, not far from the 50--75\% determined by
Le~F\`evre~et~al.~(2013); the other 459 have a photometric redshift
that is below $z=2$ and are excluded by the dataset.

In the end we include in our final database 3730 objects with $m_i<25$
for which we have measured a redshift and assigned a spectroscopic
flag from 1 to 9. Of them, 3650 are primary targets, and in addition
we have 80 secondary objects with $m_i<25$. Moreover, 231 objects with
a photometric redshift in the range $2<z<6$ and $m_i<25$ have been
targeted by spectroscopy, but no spectroscopic redshift could be
measured (they are identified by the spectroscopic flag=0). In the
next sections, we will take into account their possible contribution
to the evolution of the fraction of the \lae~emitters.

In order to allow a fair comparison with other works in the
literature, we define as strong \lae~emitters all the galaxies with a
rest--frame equivalent width of \lae~in excess of 25~\AA. In the end,
430 of the 3961 galaxies ($\sim11$\%) meet this definition.

The details about the number of objects for each flag class, as a
function of the presence of strong \lae~emission, can be found in
Table~1. The large majority of the galaxies used in this study has a
spectroscopic redshift with very high reliability: in fact, 1438
objects (36\% of the total) have a spectroscopic flag 2, meaning that
they are right in 75--85\% of the cases (Le~F\`evre~et~al.~2013); 1593
objects (42\% of the total) have a spectroscopic flag 3 or 4, that are
proven to be right in more than 95\% of the cases, 601 (15\% of the
total) are the objects with spectroscopic quality 1, but for which the
spectro-z differs less than 10\% from the photometric one and 98
objects ($\sim$2\% of the total) have a spectroscopic flag 9, meaning
that only one feature, in their case \lae, has been identified in the
spectrum, and for which about 80\% are proven to be right (Le
F\`evre~et~al.~2014). Finally, 231 objects ($\sim$6\% of the total)
have spectroscopic flag 0, meaning that a spectroscopic redshift could
not be assigned.

From Table~1, it is evident that the vast majority of objects with
strong \lae~(EW$_0>25$~\AA) have been assigned a quality flag of 3 or
4: this is not surprising, and it reflects a tendency by the redshift
measurers to assign an higher flag when the spectrum has \lae~in
emission. We note as well that not all the galaxies with flag 9 are
strong \lae~emitters, although all of them, of course, have \lae~in
emission (it is the only spectral feature identified in their
spectrum): only in $\sim40$\% of the cases is the emission strong
enough to pass the equivalent width treshold of 25~\AA.

We show in Figure~\ref{fuv} the absolute magnitude in the Far
Ultra--Violet as a function of redshift for the 3730 galaxies in the
selected sample. We compare the distribution of our galaxies with the
evolution of M$^*_{FUV}$ as derived by fitting the values for
M$^*_{FUV}$ compiled by Hathi~et~al.~(2010). In more detail,
Hathi~et~al.~(2010) derive the F$_{UV}$ luminosity function of
star--forming galaxies at $z\sim$2--3, constraining its slope and
characteristic magnitude, and compare their values with other in
the literature between $z\sim0$ and $z\sim8$. With the aim of deriving an
evolving M$^*_{FUV}$ as a function of the redshift, we took the values
published by Arnouts~et~al.~(2005) at $0<z<3$, Hathi~et~al.~(2010) at
$2<z<3$, Reddy~\&~Steidel~(2009) at $z\sim3$, Ly~et~al.~(2009) at
$z\sim2$, Bouwens~et~al.~(2007) at $z\sim4, 5, 6$
Sawicki~et~al.~(2006) at $z\sim4$ and Mc~Lure~et~al.~(2009) at
$z\sim5, 6$ and we fitted a parabola to them. In particular, we get
this best-fit:
\begin{equation}
M^*(z)=-18.56-1.37\times z+0.18\times z^2.
\end{equation}

We report this best fit on Fig.~\ref{fuv}, together with the curve
corresponding to $M^*_{FUV}+1$: we can see that the data sample quite
well the FUV luminosities brighter than M$^*$ up to redshift
$z\sim5$. Similarly, we probe the luminosity down to one magnitude
fainter than $M^*_{FUV}$ up to redshift $z\sim3.5$.  We also note that
at $z>5$, where the \lae~line and the \lae~forest absorptions by the
IGM enter the $i'-$band, we only detect the brightest UV galaxies,
while we completely miss galaxies around $M^*_{FUV}$. In the remaining
of the paper, we will be cautious to include galaxies at $z>5$ in our
analysis, and where we will do so, we will discuss the consequences.

For the analysis that we present in the following sections we build
two volume limited samples: the bright one, that contains all galaxies
brighter than $M^*_{FUV}$ at redshift $2<z<6$; and the faint one, that
contains galaxies with $M_{FUV}<M^*_{FUV}<M_{FUV}+1$, limited at
$z<3.5$. This approach is slightly different than the one used in
similar studies in the literature: Stark~et~al.~(2010, 2011) and
Mallery~et~al.~(2012), for example, rather use fixed intervals of
absolute magnitudes at all redshift. However, we prefer here to
account for the evolution of the characteristic luminosity of
star--forming galaxies, comparing at different redshifts galaxies that
are in the same evolutionary state.

\section{The distribution of the rest-frame EW of \lae}
We show in Fig~\ref{ew_dist} the distribution of the rest--frame EW of
\lae~in four redshift bins, for the bright and faint samples
separately. Positive EW indicate that \lae~is in emission, and
negative EW indicate that the line is in absorption.

Although we measured the equivalent width of \lae~for all the 3730
objects with a measured spectroscopic redshift (all the galaxies with
spectroscopic flag 2, 3, 4 and 9, and also the objects with flag 1 for
which the spectroscopic redshift differs less than 10\% from the
photometric one), this figure includes only the 3204 objects in the
bright and faint volume limited samples. These are the largest volume
limited samples of UV selected galaxies with almost full spectroscopic
information ever collected in the literature, and they allow us to
constrain the EW distribution of the \lae~line from star--forming
objects with strong \lae~in absorption compared to those with strong
\lae~in emission.  It can be seen that the shape of the distribution
is similar at all redshifts: it is lognormal and it extends from
-50\AA~ to 200~\AA, with the peak at EW$_0$=0 at all redshift and for
all luminosities.
\begin{figure*}
  \centering \includegraphics[width=\textwidth]{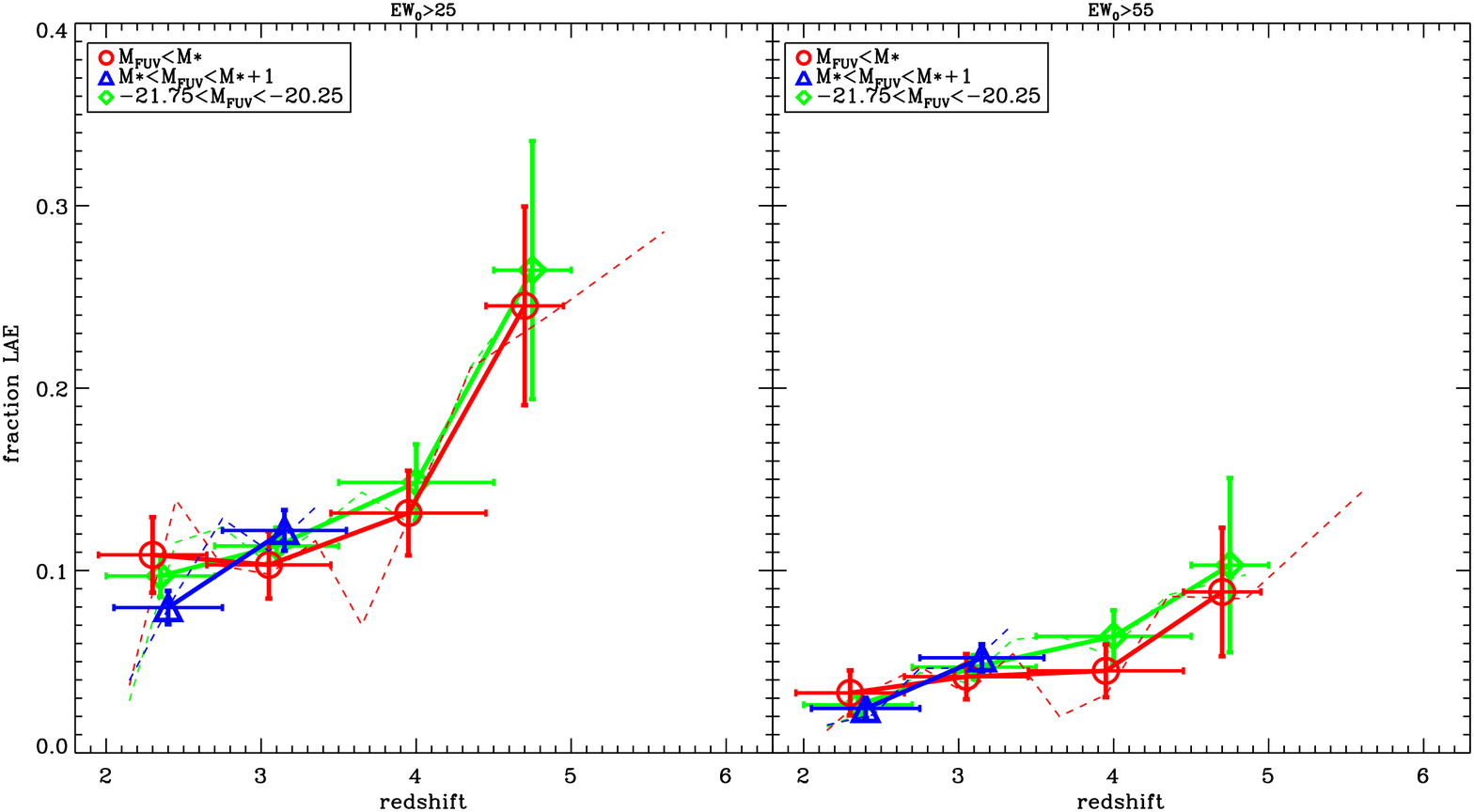}
  \caption{{\it Left panel:} Our best estimate of the fraction of
    galaxies with EW$_0$(\lae)$>25$~\AA, as a function of the
    redshift, for three intervals of far-UV absolute magnitudes: faint
    objects ($M^*<M_{FUV}<M^*+1$) are shown in blue; bright objects
    ($M_{FUV}<M^*$) are shown in red; objects with
    $-21.75<M_{FUV}<-20.25$ are shown in green. The fiducial values,
    shown by the continuous thick lines, include all the galaxies with
    spectroscopic flag 2, 3, 4 and 9, and also all the galaxies with a
    spectroscopic flag 1 and a spectroscopic redshift that differs
    less than 10\% from the photometric one. The dashed lighter lines
    show a finer binning in redshift. {\it Right panel:} same as
    left panel, but for galaxies with EW$_0$(\lae)$>55$~\AA}
  \label{frac}%
\end{figure*}

\begin{figure*}
  \centering
  \includegraphics[width=\textwidth]{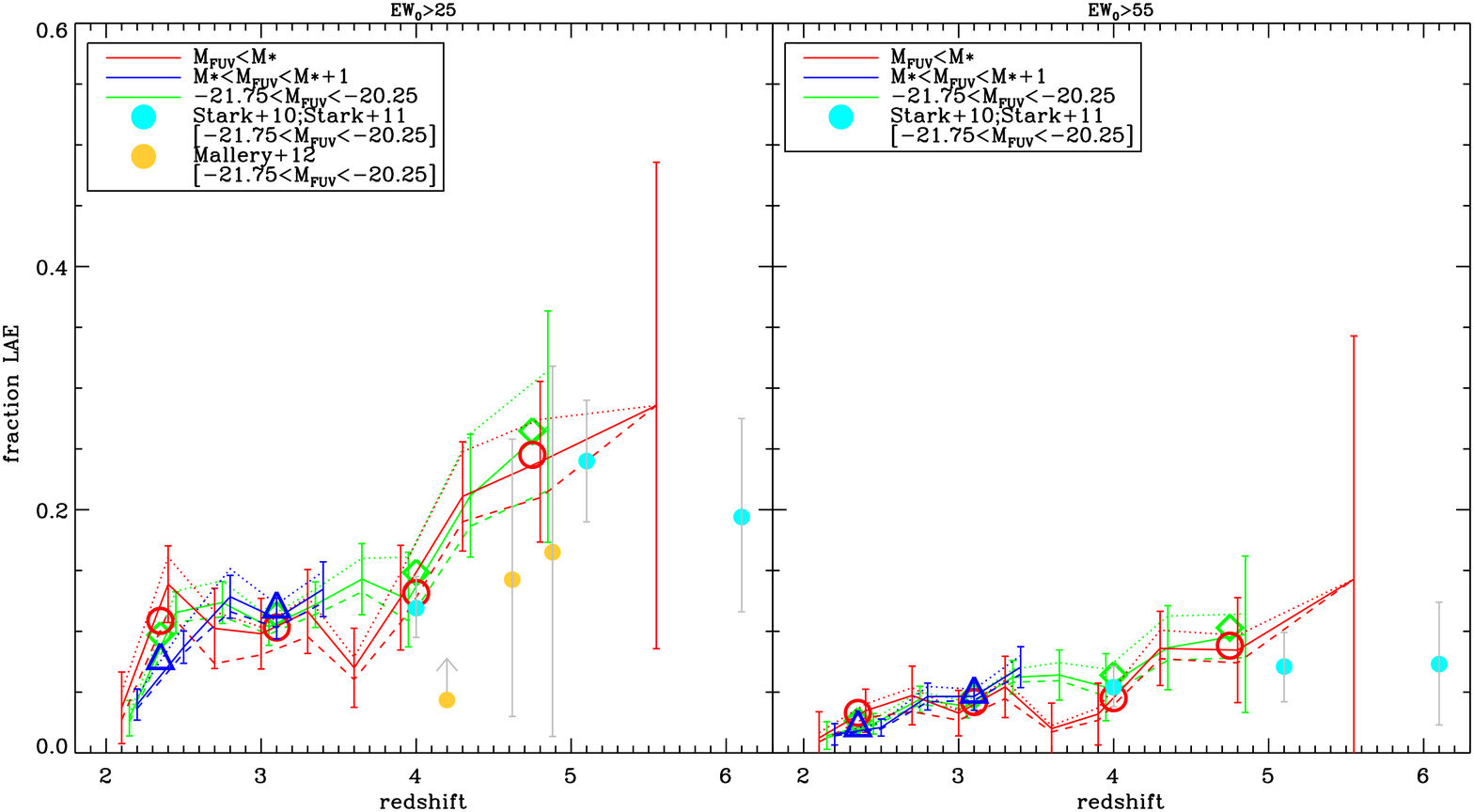}
  \caption{Same as Figure~\ref{frac}, but with a finer binning in
    redshift, and showing the effect of including galaxies with flags
    0 and 1. The left panel shows the case when the galaxies with
    EW$_0$(\lae)$>25$~\AA~are considered as emitters, and the right
    panel when the threshold is fixed at EW$_0$(\lae)$>55$~\AA.  The
    fiducial values, shown by the continuous thick lines, include all
    the galaxies with spectroscopic flag 2, 3, 4 and 9, and also all
    the galaxies with a spectroscopic flag 1 and a spectroscopic
    redshift that differs less than 10\% from the photometric one. The
    dotted line shows the case when only flags 2, 3, 4, and 9 are
    considered; the dashed line is the same as the fiducial case, but
    the galaxies with no spectroscopic redshift (flag=0) are also
    included, with the redshift fixed to the photometric one. The red
    curves are for the bright volume limited sample; the blue ones are
    for the faint one; the green ones are for galaxies with
    $-21.75<m_{FUV}<-20.25$. For clarity, the error bars are shown
    only for the continuous curves. The cyan points are from
    Stark~et~al.~(2010; 2011), the yellow ones from
    Mallery~et~al.~(2012). The red circles, blue triangles and green
    lozenges show the coarser binning in redshift adopted in
    Figure~\ref{frac}.}
  \label{fracb}%
\end{figure*}

In the first two redshift bins, $2<z<2.7$ and $2.7<z<3.5$, we can
compare the EW distributions of the bright and faint sample, and we
can see that they are quite similar. However, the extension of the
tail of objects with large \ew~evolves fast with redshift: while at
$2<z<2.7$ 11\% (7\%) of the bright (faint) galaxies have \ew$>25$\AA,
that fraction increses to $\sim15\%$ (12\%) at $2.7<z<4$ and to 25\%
at z$\sim$5. Similarly, we observe an evolution with redshift of the
upper \ew~threshold which contains 80\% of the sources: at $2<z<3.5$
the threshold is around 10--12\AA~(for galaxies in both the bright and
faint samples), at $3.5<z<4.5$ it evolves to $\sim$18\AA~and at
$4.5<z<6$ it moves to $\sim$30\AA.

We note as well that the only 13 galaxies in the whole sample have
$EW_0>150$\AA~ (the highest value is $EW_0=278.2$ at $z=2.5661$). So
extreme \ew~can not be easily produced by star formation with a
Salpeter IMF, but must have a top-heavy IMF, a very young age $<10^7$
yr and/or a very low metallicity (Schaerer~2003).

\section{The evolution of the fraction of strong \lae~emitters among star--forming galaxies at $2<z<6$}

We present in Figure~\ref{frac} the evolution with the redshift of the
fraction of star--forming galaxies that have an equivalent width of
\ew$>25$~\AA~(left panel) and \ew$>55$~\AA~(right panel), for the
bright sample ($M_{FUV}<M*$) and for the faint one
($M^*<M_{FUV}<M^*+1$) separately. In both panels we show the same
fraction for galaxies with $-21.75<m_{FUV}<-20.25$, for consistency
with previous studies (Stark~et~al.~2010; Stark~et~al.~2011;
Mallery~et~al.~2012). As we showed in Figure~\ref{fuv}, while the
bright sample ($M_{FUV}<M*$) is well represented up to $z\sim6$, the
faint one is represented only up to $z\sim3.5$: in fact, the cut in
observed magnitude at $m_i<25$, that we apply to be sure that the
continuum is detected in spectroscopy with a S/N high enough to detect
possible UV absorption features, basically prevents us by construction
from having faint galaxies in our sample beyond $z\sim3.5$.

Our fiducial case is obtained when we include all objects with
spectroscopic flag 2, 3, 4 and 9, and we also add the ``good'' flag 1
(those objects for which the spectroscopic and photometric redshifts
differ by less than 10\%) to the distribution. However, it is possible
that this combination slightly overestimates the true fraction, as we
know that 231 objects with photometric redshift $2<z<6$ have been
observed in spectroscopy, but for them a spectroscopic redshift could
not be assigned. So, it is possible that a fraction of them are
actually at $2<z<6$, and since no \lae~is present in the whole
observed spectral range, they will decrease the fraction of strong
\lae~emitters by a given amount. We discuss in Fig~\ref{fracb} the
effect on the fraction of emitters of the choice of including objects
with spectroscopic flag 0 and 1, that is quite minimal.

In Figure~\ref{frac}, for the three ranges of UV luminosities, we also
show the fractions obtained on a finer redshift grid
($\Delta~z\sim0.3$) and on a coarser grid, that highlights the general
trend and smooth out variations due to cosmic variance. The fine grid
extends on the whole $2<z<6$ range for the bright sample: however,
only the highest redshift bin contains galaxies at $5<z<6$ and might
be affected by the detection bias due to the \lae~line entering the
$i'-$band at that redshift. In the case of the coarser grid we limited
the analysis to the galaxies at $z<5$, so to be sure that the results
are not dependent on that effect.

For the bright sample, the evolution of the fraction of emitters with
\ew$>25$\AA, shown in the left panel of Fig.~\ref{frac}, is
characterized by a very modest increase in the fraction of \lae~
emitters between redshift $z\sim2$ and $z\sim4$ (from $\sim10$~\% at
$z\sim2$ to $\sim15$~\% at $z\sim4$), and then by a faster increase
above $z\sim4$ (the fraction reaches $\sim25$~\% at $z\sim5$ and
$\sim30$~\% at $z\sim5.5$). A very similar trend is observed when
galaxies with $-21.75<m_{FUV}<-20.25$ are considered. If we then
analyze the faint sample, and we compare it with the bright one, we
find that the overall fraction of objects with \ew$>$25\AA~(or
\ew$>$55\AA) is similar to that of the bright sample between $z\sim2$
and $z\sim3.5$, but the evolution between $z\sim2.3$ and $z\sim3$ is
much faster for the faint sample. This is in apparent disagreement
with the results by Stark~et~al.~(2010; 2011), who found both a higher
fraction of \lae~emitters and a steeper evolution of this fraction
among faint UV galaxies ($-20.25<M_{FUV}<-18.75$) than among bright UV
galaxies ($-21.75<M_{FUV}<-20.25$). However, we note that the range
of UV luminosities probed by our study is narrower than the one probed
by Stark~et~al.~(2010; 2011).

The right panel of Figure~\ref{frac} shows the effect on the fraction
of \lae~emitters of changing the EW threshold from 25~\AA~to 55~\AA. It
can be seen that, as expected, the fraction drastically decreases at
all redshifts. However, the general trend observed in the left panel
of Fig.~\ref{frac} is preserved: we observe that the fraction remains
around 3-4\% with a slight increase in $2<z<4$, then it increases
faster between $z\sim4$ and $z\sim5$ rising from 5\% to 12\%.

An important point that needs to be stressed again here is that our
selection criteria are completely independent of the presence and
strength of the \lae~emission up to $z\sim5$. This selection ensures
that there are no biases in the determination of this fraction over
the range $2<z<5$: if, for some reason, our selection is less complete
in a given redshift range, it will be homogeneously incomplete for
galaxies with and without \lae, and thus the result shown in this
Section will remain robust.

We report again the evolution of the fraction of \lae~emitters with
$EW_0>25$\AA~ and $EW_0>55$\AA~ in Fig~\ref{fracb}, where we simply
highlight the results on the finer redshift grid and we show the
effect of including objects with spectroscopic flag 0 and 1 in the
analysis. In this Figure, as in Fig.~\ref{frac}, we consider our
fiducial case the one including all the ``good'' flag 1 (objects with
a spectroscopic flag 1, for which the spectroscopic redshift and the
photometric one differ by less than 10\%), together with flags 2, 3, 4,
and 9. If objects with spectroscopic flag 1 are excluded, and only
flags 2, 3, 4, and 9 are considered, the fraction of emitters increases
by $\sim$2\%, with respect to the fiducial value, at all redshifts and
for all UV luminosities. This is quite obvious: since basically all
the strong \lae~emitters have a spectroscopic flag 2, 3, 4 and 9 (see
Table~1), this set of flags maximizes the fraction.  On the other
hand, if the objects with flag 0 are also considered, together with
good flag 1 and all the flags 2, 3, 4, and 9, the fraction decreases by
$\sim$2\% with respect to the fiducial case. This effect is also easy
to understand: objects with flag 0 are all non-emitters, because if an
emission line had been identified they would have been assigned a
redshift and a flag, and thus their net effect is to decrease the
fraction. Although we can not know for sure how many of these objects
with no spectroscopic redshift are indeed at the photometric redshift,
their effect is almost negligible: for both Figures \ref{frac} and
\ref{fracb}, the effect of considering flags 2, 3, 4, and 9 or of
including good flag 1 and flag 0 is always below a few percent.

We also note that our values are in good agreement with those
published by Stark~et~al.~(2010; 2011) and that are based on a
completely different method that uses LBG technique to photometrically
identify high--redshift galaxies (at $z\sim$4, 5 and 6) that are then
observed in spectroscopy to look for strong \lae~emission. Our values
are slightly higher than those by Mallery~et~al.~(2012) although
still compatible within the error bars.

\begin{figure*}
  \centering
  \includegraphics[width=.9\textwidth]{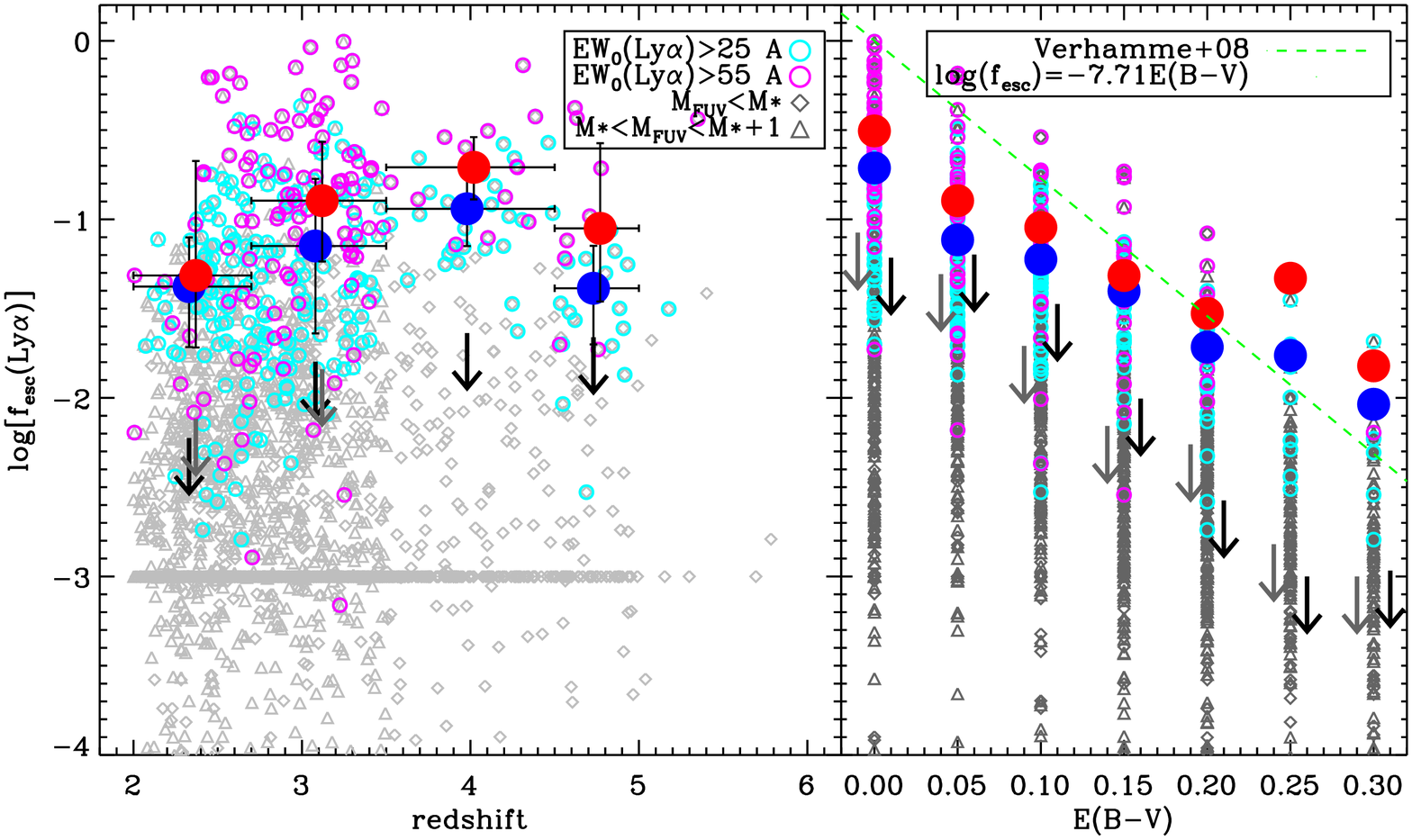}
  \caption{{\it Left panel: }\lae~escape fraction as a function of
    redshift for the bright (gray diamonds) and faint (gray triangles)
    volume limited samples. Strong \lae~emitters with $EW_0>25$\AA~
    and $EW_0>55$\AA~ are shown with cyan and magenta empty circles,
    respectively. Objects with formally negative equivalent width of
    \lae, corresponding to negative \lae~luminosity, are set here to
    $\log[f_{esc}(Ly\alpha)]=-3$. The big red and blue circles
    indicate the median escape fraction for the galaxies with
    $EW_0[Ly\alpha]>55$\AA~ and $EW_0[Ly\alpha]>25$~\AA,
    respectively. The black (gray) arrows indicate the \esc~below
    which 80\% of the bright (faint) objects lie. {\it Right panel:}
    \lae~escape fraction as a function of the E(B-V). The symbols are
    the same than in the left panel. The green dashed line shows the
    prediction by Verhamme~et~al.~(2006).}
  \label{escape}%
\end{figure*}

\section{\lae~escape fraction: driver of the \lae~fraction
evolution?}  

The escape fraction of \lae~photons $f_{esc}(Ly\alpha)$ is defined as
the fraction of the \lae~photons that are produced within a given
galaxy and that actually escape from the galaxy itself. Given the
intrinsic resonant nature of the \lae~photons, it is thought to be
dependent on the dust content, geometry of the inter--stellar gas
(ISM) and relative kinematics of the ISM and stars. Atek~et~al.~(2014)
and Hayes~et~al.~(2014), studying local samples of \lae~emitters, they
found a correlation between \esc~and E(B-V), with the escape fraction
being larger on average in galaxies with low dust
content. Kornei~et~al.~(2010) and Mallery~et~al.~(2012), although with
smaller samples than the one we use here, found a similar correlation
at $z\sim3$ and at $3.5<z<6$, respectively.

The \lae~escape fraction is usually determined by comparing the \lae~
luminosity with the dust-corrected $H\alpha$ luminosity, once a
recombination regime has been chosen. The $H\alpha$ line in fact is
not resonant and it is only attenuated by dust. However, for most of
the redshift range of this study $H\alpha$ is redshifted even beyond
the reach of near-infrared spectrographs.

An alternative method exploits the expected correlations between
intrinsic \lae~luminosity, $H\alpha$ luminosity and the SFR of the
galaxy. In particular, we assume that

\begin{equation}
f_{esc}(Ly\alpha)=L_{Ly\alpha,obs}/L_{Ly\alpha,int}=SFR(Ly\alpha)/SFR(SED),
\end{equation}

\noindent where $L_{Ly\alpha,obs}$ and $L_{Ly\alpha,int}$ are the observed and
intrinsic \lae~ luminosities, respectively; SFR($Ly\alpha$) and
SFR(SED) are the SFR obtained from the observed \lae~ luminosity and
the total SFR, respectively. Using Kennicutt~(1998) prescription to
convert $L_{Ly\alpha,int}$ into $SFR(Ly\alpha)$

\begin{equation}\label{kenn}
SFR(Ly\alpha)=L_{Ly\alpha}/(1.1\times10^{42}),
\end{equation}

\noindent we finally get

\begin{equation}
f_{esc}(Ly\alpha)=SFR(Ly\alpha)/SFR(SED)=\frac{L_{Ly\alpha}/(1.1\times10^{42})}{SFR_{SED}}.
\end{equation}

\noindent We note that Equation~\ref{kenn} assumes the case B
recombination regime (Brocklehurst~1971), that predicts an intrinsic
ratio $L_{Ly\alpha,int}/L_{H\alpha,int}=8.7$.

We stress here that the SFR inferred from fitting
Bruzual~\&~Charlot~(2003) models to the SED of galaxies give only a
crude estimate of the star formation rate and of the dust content of
galaxies. This is expecially true in the redshift regime probed by
VUDS, that is so far poorly explored, and for which independent
estimates of the SFR from different methods are scarce. However, the
SFR inferred from SED fitting are believed to be on average correct
within a factor of 3 (Mostek~et~al.~2012; Utomo~et~al.~2014), and thus
we choose to use these to obtain at least a crude estimation of the
\lae~escape fraction.

We plot in Figure~\ref{escape} the escape fraction $f_{esc}(Ly\alpha)$
as a function of the redshift and of dust reddening E(B-V) for the
galaxies in the bright and faint volume limited samples together. We
tried to separate the two samples, to check for differences among the
two them, but we did not find any, so we decided to show them
together. For the galaxies with \lae~in absorption (i.e.,
$EW_0(Ly\alpha)<0$, 1628 galaxies) we artificially set \esc~to
$10^{-3}$. For galaxies with \ew$>0$ (1576 galaxies), the \lae~escape
fraction ranges from 10$^{-4}$ to 1. We calculate as well the median
escape fraction in bins of redshift, using the same coarse grid used
for Fig.~\ref{frac}, and limiting the highest redshift bin to $z=5$,
to avoid possible detection biases affecting our selection at higher
$z.$

It is clear from this figure that at each redshift and for each E(B-V)
the strong \lae~emitters (with $EW_0(Ly\alpha)>25$\AA~ or
$EW_0(Ly\alpha)>55$\AA) are the (rare) galaxies with the highest \lae~
escape fraction. In more details, 80\% of the galaxies with escape
fraction \esc$>10$\% have \ew$>55$\AA, and 70\% of the galaxies with
\esc$>3$\% have \ew$>25$\AA. The median escape fraction for galaxies
$EW_0(Ly\alpha)>25$\AA~ is around 8\% overall, evolving from 3\% at
$z\sim2.3$ to 8\% at $z\sim3$ to 12\% at $z\sim4$. The median escape
fraction for galaxies $EW_0(Ly\alpha)>55$\AA~ is of course higher,
evolving from 5\% at $z\sim2.3$ to 12\% at $z\sim3$ to 20\% at
$z\sim4$. For both threshold we observe a decrease in the median
escape fraction between $z\sim4$ and $z\sim5$, which is probably due
to the limited amount of data.

If we then consider the whole population in our sample, and we put
together the bright and faint volume limited samples, we find that
formally the median escape fraction is zero at all redshifts. In fact,
the objects with \lae~in absorption (that have $f_{esc}(Ly\alpha)$
fixed to $10^{-3}$) are the majority, at all $z$, forcing the median
\esc~to zero. For this reason, we find more useful to show the
evolution of the \esc~below which 80\% of the galaxies, at each
redshift, lie (arrows in Fig.~\ref{escape}). Indeed, this threshold
evolves from 1\% at $2<z<2.7$ to 1.5\% at $2.7<z<3.5$ to 2\% at
$3.5<z<5,$ with not much difference between the bright and the faint
samples.

The comparison of the $f_{esc}(Ly\alpha)$ with the E(B-V) is also
interesting. From the right panel of Fig.~\ref{escape} we can see that
the E(B-V) anti-correlates with $f_{esc}(Ly\alpha)$: for objects with
high E(B-V) the median vaue of $f_{esc}(Ly\alpha)$ is low (and
vice versa). This is in qualitative agreement with the results by
Hayes~et~al.~(2014) and Atek~et~al.~(2014) in the local Universe, and
with Kornei~et~al.~(2010) and Mallery~et~al.~(2012) at
high-$z$. Moreover, the median values for the galaxies with
\ew$>25$\AA~and \ew$>55$\AA~correlates with the E(B-V) similarly to
the prediction by Verhamme~et~al.~(2006), although our data are better
fitted by a flatter slope ($\sim$-5 in comparison with -7.71 predicted
by Verhamme~et~al.~(2006). However, while galaxies with high E(B-V)
never show large $f_{esc}(Ly\alpha)$, the contrary is not true: when
E(B-V) is low we observe a broad range of \lae~escape fractions,
ranging from $10^{-3}$ to 1. This implies that the dust content alone
can not be the only factor to regulate $f_{esc}(Ly\alpha)$, at least
for galaxies with the UV luminosities similar to the ones probed in
this paper. A possibility is that in these objects \lae~ photons are
scattered at large distances, instead of being absorbed by dust,
similarly to the \lae~haloes presented by Steidel~et~al.~2011 and
Momose~et~al.~2014. We intend to test this hypothesis by stacking 2-d
spectra of galaxies with and without \lae~emission in a forthcoming
paper.

\section{Summary, discussion and conclusions}
In this paper we used the unique VUDS dataset to build an unbiased and
controlled sample of star--forming galaxies at $2<z<6$, selected
according to the photometric redshifts determined using the overall
SED of the galaxies. This selection is complementary to the classical
LBG technique, resulting in more complete and less contaminated
samples of galaxies at high--$z$. For the purpose of this paper, even
more imporant is that the combination of the selections we use
are independent of the presence of \lae~in emission, at least up to
$z\sim5$: whatever incompleteness could affect our sample, it would
affect galaxies with and without \lae~in the same way.

The sample is limited at $m_i<25$, ensuring that the continuum is
detected with S/N$\sim10$ per resolution element: this allows an
accurate determination of the spectroscopic redshift through the
identification of UV absorption features even for galaxies without
\lae~in emission.

We split this sample in two volume limited samples, using a far-UV
luminosity cut that is evolving with redshift, following the observed
evolution of $M^*_{FUV}$ (Hathi~et~al.~2010): the bright sample
include objects that at each redshift are brighter than $M^*_{FUV}$;
the faint one include objects with $M^*<M_{FUV}<M^*+1$.

We use these two samples to constrain the distribution of the EW of
\lae~of star--forming galaxies, that spans from objects with \lae~in
absorption to objects with \lae~in emission. We find that $\sim80$\%
of the star--forming galaxies in our sample have a \lae~equivalent
width $EW_0(Ly\alpha)<15$\AA.

We use our sample to constrain the evolution of the fraction of strong
\lae~emitters among star--forming galaxies at $2<z<6$. We showed in
Section~4 that the fraction of strong \lae~emitters with
$EW_0(Ly\alpha)>25$\AA~ and $EW_0(Ly\alpha)>55$\AA~ monothonically
increases with redshift, approximately at the same rate for the two EW
thresholds. The evolution is characterized by a slower phase between
$z\sim2$ and $z\sim4$, and by a faster evolution between $z\sim4$ and
$z\sim5.5$. We see no difference, at $2<z<3.5$ where both samples are
well represented, between the fraction of strong emitters in the
bright and faint volume limited samples. This is partly in
contraddiction with results by Stark~et~al.~(2010; 2011), who found
that the fraction is higher, and the rate of evolution with redshift
faster, for UV faint galaxies at $4<z<6$. However, this might be due
to the narrower range of UV luminosity probed by our work compared to
the one probed by Stark~et~al.~(2010; 2011).

Moreover, slicing our sample with the same UV luminosity limits used
by Stark ($-21.75<M_{FUV}<-20.25$) we see that the evolution of the
fraction of strong \lae~emitters (for both $EW_0(Ly\alpha)>25$\AA~ and
$EW_0(Ly\alpha)>55$\AA) is in very good agreement with the values by
Stark~et~al.~(2010; 2011), despite the different sample selection
methods and available spectroscopy. This is a very important result,
placing on firmer grounds the measures of the fraction of
star--forming galaxies with \lae~in emission. In fact, their sample is
LBG based and only the objects with strong \lae~emission are
spectroscopically confirmed.  In our case, on the other hand, we
stress that all the galaxies, with and without \lae, have a
spectroscopic redshift.

Finally, in Section~5, we have explored the possibility that the
evolution of the fraction of strong \lae~emitters is primarly due to a
change in the escape fraction of \lae~photons. We have found that, as
expected, the strong \lae~emitters are the objects for which \esc~is
the largest. We find as well that the median \esc~for the \lae~
emitters (with not much difference between objects with
\ew$>25$\AA~and with \ew$>55$\AA) evolves from $\sim$5\% at $z\sim2.5$
to $\sim$20\% at $z\sim5$.  If we try to estimate the median escape
fraction for the whole population, we find that it is formally zero at
all redshifts, since the majority of the galaxies in our sample have
\lae~in absorption, and 80\% of our galaxies have \esc$<1$\%. If we
estimate at each redshift the \esc~value below which 80\% of the
galaxies lie, we find that this value evolves from 1 to 2\% between
$z\sim2$ and $z\sim5$. It is interesting to compare these findings
with Hayes~et~al.~(2011), who integrated the \lae~and UV luminosity
functions from $z\sim0$ to $z\sim8$ and then compared the two to
estimate the average \esc~of the Universe at those redshifts.
According to Hayes~et~al.~(2011) the average escape fraction is around
5\% at $z\sim2$ and 20\% at $z\sim5$, values that are much higher than
those we obtain for our sample. This implies that for the galaxies
with UV luminosities that we sample in this paper ($M_{FUV}<M^*$ at
$2<z<6$ and $M^*<M_{FUV}<M^*+1$ at $2<z<3.5$) the average escape
fraction of \lae~photons is much smaller than the average escape
fraction of the Universe. In other words, the bulk of the
\lae~luminosity, at least in the redshift range $2<z<6$ that is probed
in this paper, is not coming from galaxies with the UV luminosities
that are probed in this work, but from galaxies that are much fainter
in the UV. In fact, Stark~et~al.~(2011) showed that the fraction of
strong (\ew$>$25\AA) emitters is higher in galaxies with
$-20.25<M_{FUV}<-18.75$ than in those with $-21.75<M_{FUV}<-20.25$,
implying a larger escape fraction for faint UV galaxies. This is also
in line with the results by Ando~et~al.~(2006), who found a deficiency
of strong \lae~emitters among UV bright galaxies and by
Schaerer,~de~Barros~\&~Stark~(2011), who also found that the fraction
of \lae~emitters rapidly increases among galaxies with fainter UV
luminosities, indicating that the bulk of the \lae~luminosity in the
universe comes from galaxies with $M_{FUV}>-20$.

Similarly to Kornei~et~al.~(2010) and Mallery~et~al.~(2012), we also
find that there is an anti-correlation between \esc~and the dust
content E(B-V): galaxies with low \esc~have preferentially a higher
E(B-V), and vice versa. This implies that the dust is a crucial
ingredient in setting the escape fraction of galaxies. However, we
note that galaxies with low extinction ($E(B-V)<0.05$) have a very
wide range of \lae~escape fractions, ranging from $10^{-3}$ to 1: this
means that the dust content, although important, is not the only
ingredient to regulate the fraction of \lae~photons that escape the
galaxy. In a forthcoming paper, we will further investigate the
dependence of \esc~on other quantities as stellar mass, star
formation rate and dust content, and on the evolution with redshift of
these correlations.

   \begin{acknowledgements}
We thank ESO staff for their continuous support for the VUDS survey,
particularly the Paranal staff conducting the observations and Marina
Rejkuba and the ESO user support group in Garching.  This work is
supported by funding from the European Research Council Advanced Grant
ERC-2010-AdG-268107-EARLY and by INAF Grants PRIN 2010, PRIN 2012 and
PICS 2013.  AC, OC, MT and VS acknowledge the grant MIUR PRIN
2010--2011.  DM gratefully acknowledges LAM hospitality during the
initial phases of the project.  This work is based on data products
made available at the CESAM data center, Laboratoire d'Astrophysique
de Marseille.  This work partly uses observations obtained with
MegaPrime/MegaCam, a joint project of CFHT and CEA/DAPNIA, at the
Canada-France-Hawaii Telescope (CFHT) which is operated by the
National Research Council (NRC) of Canada, the Institut National des
Sciences de l'Univers of the Centre National de la Recherche
Scientifique (CNRS) of France, and the University of Hawaii. This work
is based in part on data products produced at TERAPIX and the Canadian
Astronomy Data Centre as part of the Canada-France-Hawaii Telescope
Legacy Survey, a collaborative project of NRC and CNRS.
   \end{acknowledgements}


\begin{thebibliography}{}

   \bibitem[2006]{ando} Ando, M., Ohta, K., Iwata, I., et al., 2006,
     \apj, 645, L9

   \bibitem[2014]{Atek14} Atek, H., Kunth, D., Schaerer, D., et al.,
     2014, A\&A, 561, 89

   \bibitem[2012]{Bielby} Bielby, R., Hudelot, P., McCracken, H. J.,
     et al., 2012, A\&A, 545, 23

   \bibitem[2005]{bottini} Bottini, D., Garilli, B., Maccagni, D., et al.,
     2005, PASP, 117, 996

   \bibitem[2007]{Bouwens07} Bouwens, R.J., 2007, ApJ, 670, 928

   \bibitem[2009]{Bouwens09} Bouwens, R.J., 2009,
     ApJ, 705, 936

   \bibitem[2010]{Bouwens10} Bouwens, R.J., 2010,
     ApJL, 709, 133

   \bibitem[1971]{brock} Brocklehurst, M., 1971,
     MNRAS, 153, 471

   \bibitem[2010]{cardamone} Cardamone, C. N., van Dokkum, P. G., Urry, C. M.,
     et al., 2010, \apjs, 189, 270

   \bibitem[2014]{caruana} Caruana, J., Bunker, A. J., Wilkins, S.,
     et al., 2014, \mnras, 443, 2831

   \bibitem[1993]{charlot} Charlot, S., \& Fall, S. M., 1993, \apj, 415, 580

   \bibitem[1998]{cowie} Cowie, L. L, \& Hu, E. M., 1998,
     AJ, 115, 1319

   \bibitem[2012]{cuillandre} Cuillandre, J.-C. J., Withington, K.,
     Hudelot, P., et al., 2012, SPIE, 8448, Observatory Operations:
     Strategies, Processes and Systems IV, 84480

   \bibitem[2008]{deharveng} Deharveng, J.-M., et al., 2008,
     ApJ, 680, 1072

   \bibitem[2006]{dijkstra06} Dijkstra, M., Haiman, Z., \& Spaans,
     M., 2006, \apj, 649, 37

   \bibitem[2012]{dijkstra12} Dijkstra, M.,~\& Kramer, R., 2012,
     \mnras, 424, 1672

   \bibitem[2014]{dijkstra14} Dijkstra, M., Wyithe, S., Haiman, Z.,
     et al., 2014, \mnras, 440, 3309

   \bibitem[1985]{djorgovski} Djorgovski, S., et al., 1985,
     ApJ, 299, L1

   \bibitem[2010]{fontana} Fontana, A., Vanzella, E., Pentericci, L.,
     et al., 2010, \apj, 725, 205

   \bibitem[2006]{gawiser06} Gawiser, E., et al., 2006,
     ApJ, 642, 13

   \bibitem[2007]{gawiser07} Gawiser, E., et al., 2007,
     ApJ, 678, 278

   \bibitem[1996]{giavalisco} Giavalisco, M., Koratkar, A., Calzetti,
   D., 1996, ApJ, 466, 831

\bibitem[Giavalisco et al. (2004)]{giavalisco04} Giavalisco, M.,
  Dickinson, M., Ferguson, H. C., et al. 2004, \apj, 600, 103

   \bibitem[2007]{gronwall} Gronwall, C., et al., 2007,
     ApJ, 667, 79

   \bibitem[2012]{grogin} Grogin, N., Kocevski, D. D., Faber, S. M.,
     et al., 2012, \apjs, 197, 35

   \bibitem[2014]{guzzo} Guzzo, L., Scodeggio, M., Garilli, B., et al.,
     2014, A\&A, 566, 108

\bibitem[2010]{hayes10} Hayes, M., \"Ostlin, G., Schaerer, D.,
  et al., 2010, \nat, 464, 562

\bibitem[2011]{hayes11} Hayes, M., Schaerer, D., \"Ostlin, G., et al.,
  2011, \apj, 730, 8

\bibitem[2014]{hayes14} Hayes, M., \"Ostlin, G., Duval, F., et al.,
  2014, \apj, 782, 6

\bibitem[2010]{hathi10} Hathi, N. P., Ryan, R. E., Jr., Cohen, S. H.,
  et al., \apj, 720, 1708

\bibitem[2013]{hathi13} Hathi, N. P., Cohen, S. H., Ryan, R. E., et al.,
  2013, \apj, 765, 88

   \bibitem[2004]{hu} Hu, E. M., et al., 2004,
     AJ, 127, 563

   \bibitem[1998]{kennicutt} Kennicutt, R. C., et al., 1998,
     ApJ, 498, 541

   \bibitem[2007]{koekemoer07} Koekemoer, A. M., Aussel, H., Calzetti, D.,
     et al., 2007, \apjs, 172, 196

   \bibitem[2010]{kornei} Kornei, K. A., Shapley, A. E., Erb, D., et
     al., 2010, \apj, 711, 693

   \bibitem[1998]{kunth} Kunth, D., et al., 1998, A\&A, 334, 11

   \bibitem[2004]{LeFevre04} Le F\`evre, O., Mellier, Y., McCracken, H. J.,
     et al., 2004, A\&A, 417, 839

   \bibitem[2005]{Le Fevre} Le F\`evre, O., Vettolani, G., Garilli,
     B., et al., 2005, A\&A, 439, 845

   \bibitem[2013]{Le Fevre} Le F\`evre, Tasca, L. A. M., Cassata, P.,
     et al., 2013, submitted to A\&A

   \bibitem[2007]{Lilly} Lilly, S. J., Le F\`evre, O., Renzini, A.,
     et al., 2007, \apjs, 172, 70

   \bibitem[2003]{Lonsdale} Lonsdale, C. J., Smith, H. E., Rowan-Robinson, M.,
     et al., 2003, PASP, 115, 897

   \bibitem[2012]{mauduit} Mauduit, J.-C., Lacy, M., Farrah, D., et al., 2012,
     PASP, 124, 1135

   \bibitem[2003]{mas-hesse} Mas-Hesse, J. M., et al., 2003, 
ApJ, 598, 858

   \bibitem[2012]{mallery} Mallery, R. P., Mobasher, B., Capak, P.,
     et al., 2012, \apj, 760, 128

   \bibitem[2010]{maraston} Maraston, C., Pforr, J., Renzini, A.,
     et al., 2010, \mnras, 407, 830

   \bibitem[2003]{mccracken12} McCracken, H. J., Milvang-Jensen, B.,
     Dunlop, J., et al., 2012, A\&A, 544, 156

   \bibitem[2014]{momose} Momose, R., Ouchi, M., Nakajima, K., et al.,
     2014, \mnras, 442, 110

   \bibitem[2012]{mostek} Mostek, N., Coil, A. L., Moustakas, J.,
     et al., 2012, \apj, 746, 124

   \bibitem[2007]{murayama} Murayama, T., et al., 2007,
     ApJS, 172, 523
     
   \bibitem[2009]{nilsson} Nilsson, K. K., et al., 2009,
     A\&A, 498, 13
     
   \bibitem[2012]{ono} Ono, Y., Ouchi, M., Mobasher, B., et al.,
     2012, \apj, 744, 83

   \bibitem[2008]{ouchi} Ouchi, M., Shimasaku, K., Akiyama, M., et
     al., 2008, \apjs, 176, 301
     
   \bibitem[2011]{papovich} Papovich, C., Finkelstein, S. L.,
     Ferguson, H. C., et al, \mnras, 412, 1123
     
   \bibitem[1967]{partridge} Partridge, R. B, \& Peebles, J. E., 1967,
     \apj, 147, 868

   \bibitem[2011]{pentericci} Pentericci, L., Fontana, A., Vanzella, E., et al. 2011, \apj, 743, 132

   \bibitem[2006]{reddy} Reddy, N. A., Steidel, C. C., Erb, D. K.,
     \apj, 653, 100

   \bibitem[2012]{reddy} Reddy, N. A., Pettini, M., Steidel, C. C., 
     et al., \apj, 754, 25

   \bibitem[2007]{sanders} Sanders, D. B., Salvato, M., Aussel, H.,
     et al., 2007, \apjs, 172, 86
     
   \bibitem[2003]{schaerer} Schaerer, D., 2003,
     A\&A, 397, 527
 
   \bibitem[2011]{schaerer11} Schaerer, D., de~Barros, S., \& Stark,
     D. P., 2011, A\&A, 536, 72

   \bibitem[2005]{scodeggio} Scodeggio, M., et al., 2005,
     PASP, 117, 1284
     
   \bibitem[2007]{scoville} Scoville, N., Aussel, H., Brusa, M., et al., 2007,
     \apjs, 172, 1

   \bibitem[2003]{sha} Shapley, A., et al., 2003,
     \apj, 588, 65

   \bibitem[2012]{schenker} Schenker, M. A., Stark, D. P., Ellis, R. S., et al. 2012, \apj, 744, 179

   \bibitem[2010]{stark10} Stark, D. P., Ellis, R. S., Chiu, K., et
     al., 2010, \mnras, 408, 628

   \bibitem[2010]{stark10} Stark, D. P., Ellis, R. S., \& Ouchi,
     M. 2011, \apj, 728, 2

   \bibitem[1999]{steidel} Steidel, C. C., et al., 1999,
     ApJ, 519, 1
 
   \bibitem[2000]{steidel} Steidel, C. C., Adelberger, K. L.,
     Shapley, A. E., et al., 2000, ApJ, 532, 170
 
   \bibitem[2011]{steidel} Steidel, C. C., Bogosavljevic, M.,
     Shapley, A. E., et al., 2011, ApJ, 736, 160
 
   \bibitem[2007]{Taniguchi} Taniguchi, Y., Scoville, N., Murayama, T.,
     et al., 2007, \apjs, 172, 9
     
   \bibitem[1999]{tresse} Tresse, L., Maddox, S., Loveday, J., \&
     Singleton, C., 1999, \mnras, 310, 262
     
\bibitem[2014]{utomo} Utomo, D., Kriek, M., Labb\'e, I., et al.,
  2014, \apj, 783, 30

\bibitem[2011]{vanzella} Vanzella, E., Pentericci, L., Fontana, A., et al.,
  2011, \apj, 730, 35

\bibitem[2006]{verhamme06} Verhamme, A., Schaerer, D., \&
  Maselli, A., 2006,  A\&A, 460, 397

\bibitem[2008]{verhamme08} Verhamme, A., Schaerer, D., Atek, H., \&
  Tapken, C., 2008,  A\&A, 491, 89

\bibitem[2012]{verhamme12} Verhamme, A., Dubois, Y., Blaizot, J., et
  al., 2012, A\&A, 546, 111
     
   \end{thebibliography}
\end{document}